\documentclass[useAMS,usenatbib,usegraphicx]{mn2e}
\title{Period switching in the symbiotic star BX Mon}
\author[Elia M. Leibowitz and Liliana Formiggini]
{Elia M. Leibowitz$^{1}$\thanks{E-mail:
elia@wise.tau.ac.il}
and Liliana Formiggini$^{1}$\thanks{E-mail: lili@wise.tau.ac.il}\\
$^{1}$The Wise Observatory and the School of Physics and Astronomy, Raymond
and Beverly Sackler Faculty of Exact Sciences \\ Tel Aviv University, Tel
Aviv 69978, Israel\\}
\def\Pa{P$_{a }$}
\def\Pb{P$_{b }$}

\def\Pr{P$_{r }$}
\def\PL{P$_{l }$}
\def\P1{P$_{1 }$}
\def\P2{P$_{2 }$}
\def\P3{P$_{3 }$}
\begin{document}

\maketitle
\label{firstpage}
\begin{abstract} We report on a detailed analysis of the optical light
curve of the symbiotic system BX Mon, the data of which were gathered 
from the literature. The light curve covers the period December 1889  
March 2009, with a gap of no observations between March 1940 and 
February 1972. The light curve is characterized by strong oscillations 
of peak to peak amplitude of 2 to more than 3 magnitudes. Before the 
gap the fluctuations were modulated mainly by a period \Pa=1373$\pm$4 d. 
After the gap the dominant periodicity is \Pb=1256$\pm$16. Higher 
harmonics as well as a few beats of the two major periodicities can 
also be identified in the light curve. We identify one of the beat 
periods, \Pr=656 d,  as the sidereal rotation period of the giant 
component of the system . The period switching that took place 
during the gap in the observations was possibly associated with a 
certain cataclysmic event, hints of which may be
recognized in the LC in the first 11 years after the gap.

We suggest that the origin of the major oscillations is in periodic 
episodes of mass accretion from the M giant onto the hot component 
of the system. After the gap they are correlated with the periastron 
passage of the system, and therefore appear with the binary period. 
Before the gap the oscillations appeared with the diurnal cycle of 
an observer on the surface of the rotating M giant, whose sun is 
the hot component. The event of the period switching is possibly 
related to an intensive magnetic activity in the outer layers of 
the giant star. 
\end{abstract} 
\section{Introduction}
BX Mon was discovered on a Harvard objective  prism plate by Mayall (1940).
Its classification as a symbiotic system (SS) was based on its optical spectrum, 
showing a combination of strong hydrogen emission lines and TiO absorption bands 
of a  late-type star (Iijima 1985, Kenyon 1986, Viotti et al. 1986). 
Its identification as SS has been questioned (Allen 1982), until
medium ionization lines have been identified in IUE (International
Ultraviolet Explorer spectra)  (Michalitianos 1982). Its infrared colors
and spectral energy distribution are that of a normal M5 III star and
exclude the presence of a Mira variable (Whitelock \& Cathpole 1983, Viotti
et al. 1986, Dumm et al. 1998). No nova-like eruption event has been
recorded for this system.

BX Mon large photometric variability  had already been discovered in 1940
by Mayall (1940) and a period of 1380 d was suggested, with ephemeris
JD(max)= 2412490 d.  Its optical and ultraviolet spectrum is also strongly
variable. However, the spectroscopic
variability is hardly explained by the Mayall's period  (Iijima 1985,
Viotti et al. 1986). Iijima (1985) noted that two epochs of low excitation
states seem to occur at phases near the photometric maximum of Mayall's
ephemeris,  suggesting that the Mayall's light curve representation may
contain a mistake or that a change occurred in the variations phase.

Dumm et al. (1998) analyzed the Mayall's data and the data of the  RASNZ
(Royal Astronomical Society of New Zealand) obtained between the years 1989  
and 1995, covering less than two cycles. Two possible periodicities have been
suggested, P=1338$\pm$8 d and P=1401$\pm$8 d. The period P= 1401 d seems to 
explain the IUE spectra variations (Dumm et al. 1998) and in particular the 
UV flux attenuation as due to eclipses of the hot component by the cool one. 

Fekel et al. (2000), combining their  own radial velocity measurements and
a few old data by Dumm et al. (1998) established an orbital period of
1259  $\pm$16 d. A  period of P= 1262 $\pm$32 d was determined from the RASNZ
data analysis, that is consistent with the radial-velocity period.
More recently, Brandi et al. (2009) reanalyzed old and new radial velocity 
data and suggested the value of 1290 d for the binary period.
  
In order to better understand the nature of BX Mon variability, we have
reanalyzed  the long-term light curve (LC) of BX Mon using the Mayall's
data and the AAVSO (American  Association of  Variable Stars Observers)
ones.

Section 2 presents the data set that we analyze in this paper. Section 3
describes the analysis that we applied on the data and our main results.
In Section 4 we identify the clocks that give rise to the periodic variations 
in the LC of the system and propose a qualitative model which could 
explain how these clocks modulate the optical brightness of the system. 
Section 5 is a brief summary.
\section{THE HISTORIC LC OF  BX MON}
\subsection{The data}
Our reconstructed LC spans the time interval from December 1889 up to
March 2009, i.e. from JD=2411380 to JD=2454905,
with a large gap of more than 11000 days, from March 1940 to February 1972, 
for which we could not find any reported magnitude measurement. 
We refer to the two data sets on  the two sides of the gap as section A 
and section F+B (the meaning will be explained below).
Section A of the LC was retrieved from Mayall's (1940) figure 1, which
presents  the estimated photographic magnitudes of the star
on the Harvard plates from 1889 to 1940.
Our section F+B of the LC is taken from  the AAVSO data base of
measurements covering the years between  1972 and 2009. We have
binned the visual AAVSO data on 10 days averages.

Fig. 1 presents the LC of BX Mon showing variations, reaching  more than 3 mag
amplitude, and the large  gap in time between the two sets of observations.
The solid line in Fig. 1 is a running mean over a linearly interpolated LC
of the system, with a running window of 2810 d.
We have broken down section F+B into two distinct sections F and B and the
three sections of the LC are marked in Fig. 1.
Section F covers  the 4010 days between JD=2441365 and JD= 2445375.
In this section the average luminosity of the system over $\sim{2810}$
days was significantly larger than during any other interval of the same
length, as evidenced by the running mean curve in the figure.
Two fluctuations in the brightness of the system occur in this section that
ends with a rapid decline of the mean brightness back to the mean
magnitude that the system had in section A.  
\begin{figure}
\includegraphics[width=90mm]{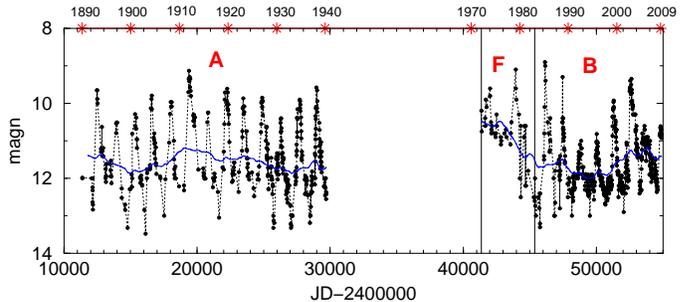}
\caption{ The historical LC of BX Mon. Sections A, B and F are explained in
the text.
The line is a running mean over a linearly interpolated LC, with a running
window
of 2810 d.}
\end{figure}      
\section {Data Analysis}
\subsection{Power spectrum}
Fig. 2 (a) is the power spectrum (PS) of the LC shown in Fig. 1, computed
according to the Lomb-Scargle prescription (Scargle 1982), and Fig. 2 (b)
shows the window function. As clearly seen, there are quite a number of 
peaks in the PS around the frequency corresponding to P=1300 d that are 
statistically significant. However, folding the LC onto any one of the 
corresponding periodicities demonstrates that none of these periods fit 
the data of the entire LC.
 
Fig. 2 (c) is the PS of section A of the LC.  Four peaks stand 
out clearly above the noise level. Their corresponding periodicities 
in days are listed in Table 1, the highest one corresponds to the 
period  \Pa=1373 d. We remark here that the period values 
in table 1  are the ones we derive by the fitting process that we 
apply as explained in  section 3.2. For completeness we list here 
the slightly different period values associated with the frequencies 
of the 4 peaks in the PS. 
They are: 1370 $\pm$10, 687 $\pm$ 2.5, 498 $\pm$1.4 and 343 $\pm$1  days. 
The quoted error estimates correspond to the half widths at half maximum 
of the corresponding peak profile in the PS. 
There is hardly any doubt that the second and 4th  periodicities 
in section A are the 2nd and 4th harmonics of the major one  \Pa,
while the 3d one is an alias as identified by a "clean" routine
that we have developed and applied to the data.

Fig. 2 (d) presents the PS of section B of the LC. The 7 highest 
peaks are marked in the figure. The peak that is slightly higher than 
the 7th marked peak is shown by the "clean" routine to be an alias of the major, 
dominant periodicity in this section of the data. The periods corresponding 
to the 6 independent peaks are listed in Table 1. Again, the numbers in the 
table are the values derived by least square fitting, as explained below. 
The periods that correspond strictly to the peaks in the "clean" PS are: 
1261 $\pm$ 17, 7706 $\pm$ 623, 1067 $\pm$ 12, 1710 $\pm$ 31, 431 $\pm$ 2.0, 
461  $\pm$ 2.2 and 661 $\pm$ 4.6 d.  
\begin{figure}

\includegraphics[width=80mm]{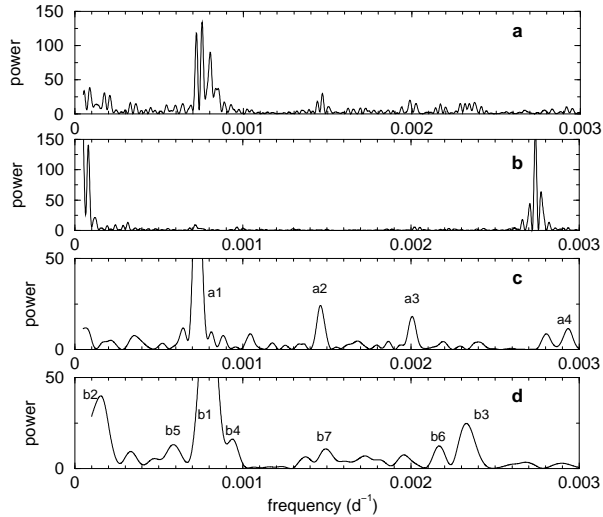}
\caption{(a) The PS of the  LC of BX Mon. (b) The window function. 
(c) The PS of section A of the LC.
(d) The PS of section B of the LC.}
\end{figure} 
\begin{table}
\caption{ Peaks in the power spectrum  } 
\begin{tabular}{@{}crll@{}} \\
    
Number &Period days & Comments & \\
\hline
Section A  & & & \\   
\\
      a1 &       1373   &  \Pa          & (1/\Pr-1/\Pb)$^{-1}$   \\
      a2 &        687   &  P$_{a}$/2    &   \\
      a3 &        498   &  alias        & \\
      a4 &        343   &  P$_{a}$/4    &   \\
\\
Section B & & &\\
\\
    b1&  1256     &  P$_{b}$  &   \\
    b2&  7370     & beat  & .5 x (1/\Pb-1/\Pa)$^{-1}$    \\
    b3&  1073     & beat &   (3/\Pb-2/\Pa)$^{-1}$\\
    b4&  1687     & beat  &  (3/\Pa-2/\Pb)$^{-1}$    \\
    b5&   431     & beat  &  (1/\Pa+2/\Pb)$^{-1}$\\
    b6&   461     & alias  & \\
    b7&   656     & P$_{r}$  & (1/\Pa+1/\Pb)$^{-1}$  \\
   
\end{tabular} 
\end{table}
\begin{figure}
\includegraphics[width=90mm]{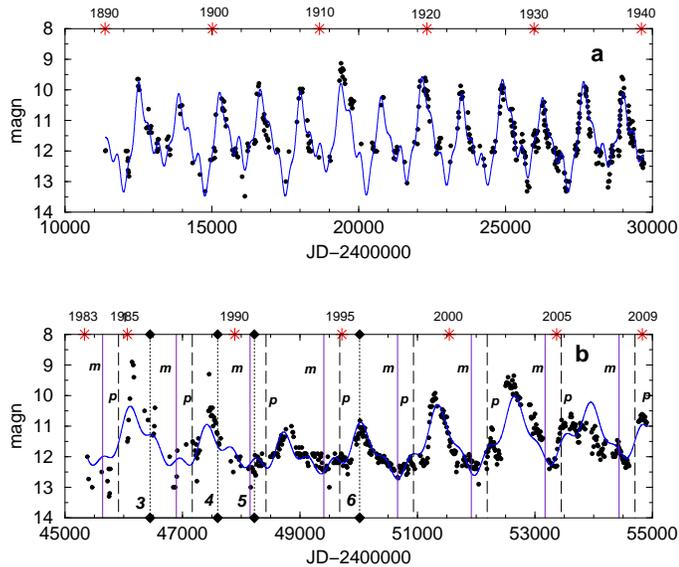}
\caption{(a) section A  of the LC of BX Mon and the fitted LC, (b) 
section B of the LC. The epochs of photometric minima are marked by m (vertical solid lines), and those of periastron by p (vertical dashed lines). Diamonds and dotted lines  mark the dates of IUE spectra}
\end{figure}
\subsection{Harmonics and Beat periods}
The major period in section B of the data is \Pb=1256 d. We identify 
it as the binary period of the  BX Mon system. 
One then finds out that the frequencies of all the other true periods 
listed in Table 1 are 
simple linear combinations of the 2  major periodicities \Pa and \Pb, 
with very small integer numbers as coefficients, as specified in 
column 3 of Table 1. We obviously can express the frequencies of all 
periods listed in Table 1 as linear combinations of \Pr~ and \Pb, rather 
than of the pair \Pa~ and \Pb.

We now assume that all observed oscillations in the LC of BX Mon are indeed these 
combinations of the two major ones \Pa~ and \Pb. We reverse the period search 
process and by least square procedure we look for the pair of periods around 
1260 and 1373, that combined with the harmonics and their appropriate beats, 
produce the best fitted multi-periodic waves to section A and section B of the 
data set. The second column of Table 1 lists the values of the two major 
periodicities \Pa~ and \Pb so obtained, and of their respective beat periods. 

With the Bootstrap technique (Efron and Tibshirani 1993) we have established a 
99\% confidence interval of 4 days around the \Pa~ period value, and of 16 days around 
the \Pb~ value.
  
Fig. 3 a) displays section A of the LC of BX Mon (dots). The solid curve is the best 
fitted wave to the data with the period \Pa~ and its 2nd and 4th harmonics. 
Fig. 3 b) depicts the data points of section B, along with the best fitted wave with 
the periods listed in Table 1. Fig. 4 a) shows section A of the data set folded onto 
the \Pa~ period. Fig. 4 b) is the folded section B data onto the period \Pb .
The ephemeris for \Pa~ is  T$_{0}$(a)= JD 2428499+1373E.

Fig. 3 b) and 4 b) show that in section B there is no well defined minimum point that 
appears with strict \Pb~ periodicity. The minimum points in each cycle of the smooth 
curve with the 6 periodicities of section B fitted to the data are also distributed 
on the time axis with only quasi \Pb periodicity. The minimum 
points of each cycle of the observed data set are scattered around this ephemeris 
with a dispersion of $\sim100$ days.
Therefore exact photometric ephemeris for the \Pb periodicity in section B are not 
well defined. Ephemeris of a Sine wave with the \Pb period that best fit the data is: 
 
T$_{0}$(b)= JD 2449445 ($\pm$ 100)+1256E
    
\subsection{SED considerations}
\subsubsection{The IUE observations}
Table 2 lists the  13 low resolution observations obtained by the IUE 
(International Ultraviolet Explorer) between 1979 January and 1995 October, retrieved 
from  the  INES Archive Data for archival spectra.

The phase listed is  computed with respect to  our ephemeris, minimum  
light JD$_{Pb}$ =2449445 ($\pm$ 100) d  (see section 3.2). 
We identify 5 epochs of UV observations. The exposure 
classification code of the observation LWP31630L shows that no
continuum spectrum was detected and we discarded this epoch. 

As mentioned in Section 1, Dumm et al. (1998) noted two episodes of strong
attenuation in the UV flux of the system as measured by the IUE telescope.
The two episodes are separated from each other by 4347 days. In trying to
interpret the events as an eclipse phase in the binary cycle of the system
Dumm et al. (1998) were forced to postulate a period of 1401 days.
 
The first IUE event, that took place at epoch 1,  falls within our 
section F of the LC. As indicated by the mean brightness of the star, at 
that time some other unknown process dominated the optical luminosity of the system, 
probably the end of the event that brought about the period switching in the 
quiescence state of the system. Therefore the timing of that event of attenuation 
of the UV flux is probably unrelated to the clocks that drive the two major 
periodicities of the system. In fact, in the same section F, the two optical 
fluctuations in the brightness of the system are also not in phase with neither 
one of the two dominant periods in the LC. 

The other IUE measurement of attenuated UV is epoch 5. There is virtually no 
trace of UV continuum emission from BX Mon in the wavelength range 100-180 nm 
at that time, see Fig 5. Unfortunately, no frame of the LWP 
camera of IUE of the same date exists in the IUE archives. 
The lack of continuum emission may indicate an eclipse of the white dwarf (WD) of 
the system by the giant star, namely, conjunction. This, however, is not 
necessarily so. 

Lack of far UV emission from the hot component of symbiotic 
stars has been recorded also when no eclipse is expected. 
An example is the IUE measurement  SWP09385LL in AG Dra taken on
June 27 1994, near  maximun light of the photometric binary cycle
(phase=.96). 
Note also that one cycle of 1256 days after the date of no short wavelength 
continuum in the IUE measurement in BX Mon falls 24 day after the date 
suggested by Brandi et al. (2009) as the spectroscopic conjunction of the system. 
With the periodicity suggested by these authors the mismatch is 58 days. 
The cycle following the IUE measurement falls also 38 days after the ephemeris time 
that we mentioned above as obtained by fitting a Sine wave of the \Pb periodicity 
to the entire data set of section B.     
\begin{figure}
\includegraphics[width=60mm]{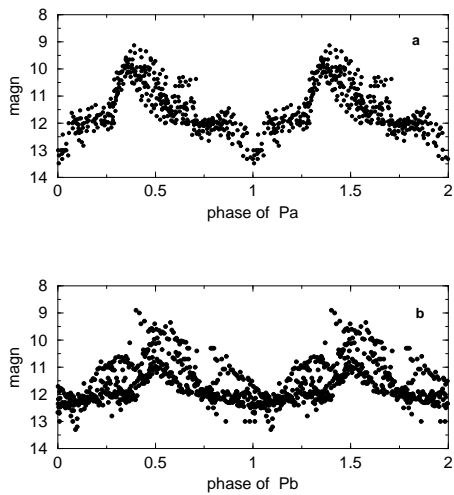}
\caption{(a) section A  of the LC of BX Mon folded on P$_{a}$=1373 d,  
(b) section B of the LC folded on P$_{b}$=1256 d.}
\end{figure}
\begin{table}
\caption{IUE spectra of BX Mon   } 
\begin{tabular}{@{}crllll@{}} \\
    
Image  &  Date  &   JD  &   exp time &    ph2 &  epoch\\
      &         &       &     sec             & \\
\hline 
SWP03832L & 1979-01-06&  24 43880 & 4619&   .57    &    1)\\
LWR03408L & 1979-01-06&  24 43880 & 2400&          &  \\
SWP06344L&  1979-09-01&  24 44118 &  3600 &  .76   &    2)\\
LWR05479L&  1979-09-01&  24 44118 &  3600&         & \\
SWP27797L&  1986-02-26 & 24 46488&   3600&   .65  &   3)\\
LWP07724L&  1986-02-26&  24 46488&   2400&          &    \\
SWP35767L&  1989-03-14&  24 47600&   5400&    .53  &     4)\\
LWP15196L&  1989-03-14&  24 47600&   1800&        &    \\ 
SWP40243L&  1990-12-01&  24 48227&   7200&   .03&       5)\\
SWP56128L&  1995-10-27&  24 50017&  14400&      &    \\
LWP31630L&  1995-10-27&  24 50017&   3600&         &    \\
SWP56132L&  1995-10-29&  24 50019&   6300&  .46  &     6)\\
LWP31630L&  1995-10-29&  24 50019&   3600&       &    \\
\end{tabular} 
\end{table}

\begin{figure}
\includegraphics[width=70mm]{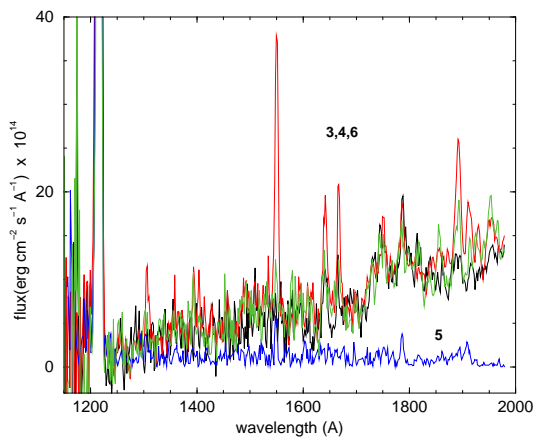}
\caption{The short wavelength spectra of observations n.3,4,6}
\end{figure} 
\subsubsection{Attempts of model fitting}
Fig. 5  shows the short wavelengths behaviour of all the IUE epochs within the time interval of our section B. 
The shape of the short wavelength continuum  
of  n.3, 4, 6 is almost the same. In observation n.5 only the emission 
lines are present while no continuum is detected. We retrieved the UBVRIc 
photometry (Munari et al. 1992) and a few JHKL 
measurements (Henden \&  Munari 2008, Phillips 2007), that  have been transformed to 
the Johnson system (Bessel 1983) and to flux, according to the standard procedure 
(Zombeck 2000, page 100). Data have been dereddened by E(B-V)=.25 (Dumm et al. 1998). 
We matched the JHKL photometry at JD=24 47963, JD=24 47996 and  UBVRcI data at JD=24 27977
with the IUE epoch 4.
The ultraviolet SED for epoch 6, at maximum light, is quite similar to that of epoch 4, 
as shown in Fig. 5. There are no photometric data that can be  matched with 
this epoch. The nearest in time UBVRcI photometry, is at JD=24 51582, a cycle later and 
at phase =.66. 

Using methods presented also by Skopal (2005), we tried to construct  a physical 
model of the various radiation sources within this stellar system, i.e. cool giant 
star, hot component, emission nebula, possible accretion disk. Such an attempt has 
been made already in the past by Kenyon \& Webbink (1984) who were unable to come 
up with an acceptable fit. 

Our attempts have also been severely hampered by the scarcity of multiwavelength 
spectral observations in this star, especially such that are performed simultaneously, 
even 27 years after the pioneering work of Kenyon \& Webbink (1984). Simultaneity is 
of course a necessary condition for an analysis of this kind to be meaningful for a 
system such as BX Mon the energy output of which is varying in time by nearly two 
orders of magnitude.
We do note, however, that in all  our  numerical attempts of model fitting, the contribution 
of the nebula to the IUE SWP camera is negligible. Therefore the SED No. 3,4, and 6 
shown on Fig. 5 are most probable representing the contribution of the hot component of the system.

Two of them  (No.  4  and 6) were made close to epochs of maximum brightness in the 
optical light curve of our Section B (see Fig. 3 b). These spectra are comparable 
to IUE spectra during active events of other symbiotic stars such as Z And and AR Pav 
(see figs 3 and 19 of Skopal 2005). This fact and the resemblance of the optical light curve
of BX Mon with those of  Z And and AG Dra (Skopal 2007, figs 2 and 8), for example, 
strengthen our claim, made in the following section, that in BX Mon  the cyclic outbursts 
of the system are events of intense mass transfer.

\section{Discussion}
Our time series analysis, in particular Fig. 3, suggests that two photometric
cycles characterize the major brightness variations of BX Mon in the last 120 years,
\Pa=1373 d and \Pb=1265 d.  The period \Pa~ dominated the first 55 years, while
\Pb~ is the major one in the last 26 years of the star history. The structure 
of the entire LC observed in section A can be well reproduced by an harmonic 
wave of the \Pa~ periodicity with its 2nd and 4th harmonics, as seen in Figure 4. 
The variability in section B is well represented by the major period \Pb and 
the 5 other periods listed in Table 1, all of which are simple linear 
combinations of \Pa~ and \Pb . Fig. 3 a) shows the sine wave of the period 
\Pa~ and its 2nd and 4th harmonics fitted to LC of section A. Fig. 3 b) 
is the fit to section B of the series of \Pb~ with  the 5 beats  periods
listed in Table 1.
 
We note that the difference in the periods between sections A and B of the LC can 
hardly be attributed to the relatively small difference in the colors of the two LCs, 
the Pg passband of the Harvard plates, and the eye response function of the AAVSO 
observations. An orbitally-related period has usually different amplitudes in 
different passbands, but never a different value. This difference must therefore 
be due to a real change in the period of the cyclic outbursts of the system that 
took place between the epoch before the gap in the observations around the middle 
of the 20th century, and the epoch  after it. 
  
\subsection{Interpretation}
\subsubsection{The rotational period of the giant}
 
Before discussing the two major periodicities of the system we comment on 
the nature of the period \Pr=656 d that is identified in the LC of section B 
of the data (see Section 3.1). We suggest that this is the sidereal rotation 
period of the giant component of the BX Mon binary system.  

Zamanov et al. (2008) measured the rotational velocity of this giant, quoting 
two possible values, 11.0$\pm$1.5  and 9.4$\pm$1.5 km/s. 
Taking their first suggested value,  adopting the 
value $139.6 R\odot$  for the radius of the M5III giant of this system 
(van Belle et al. 1999), and assuming that the equatorial plane of the giant 
is seen roughly edge on, the resulting rotational period of the giant is 
P$_{rot}$= 646 d.  This is almost identical to our number 656 d obtained 
from the Fourier analysis. Thus our interpretation of the results obtained by 
our time series analysis of the LC of the star as the rotation period of the giant, 
leads us to practically the same consequence about this period that was obtained 
by entirely independent observers and method.  
\subsubsection{The two major periodicities}
Our analysis shows that the two major periods of the system, \Pa=1373 d and 
\Pb=1256  d do not exist simultaneously in the LC of BX Mon, but rather that 
a major switch between the two took place during the gap in the record 
of the observations. Such a large change of 120 days in the photometric period 
of the system within 32 years is quite abrupt. To the best of our knowledge, such 
a discrete change from one periodicity to another, both with amplitudes that 
amounts to 80-90\% of the total optical emissivity of the star has not been 
observed so far in any other symbiotic star. We venture to say not even in 
any other star. There must have been some dramatic event or process that 
took place in the BX Mon stellar system that was the cause of this period 
switch. A likely photometric remnant of this event may be detected in the 
first  4000 days after the gap in the observations. For some 2500 days during 
that time, the mean brightness of the system was higher than at any other 
interval of the same time length. 
Also, as mentioned already in Section 3.3, the two light fluctuations recorded 
during these 4000 days are not in phase with either one of the two major 
periods of the system. 

The value of \Pb~ is almost certainly the binary period of the
system as determined spectroscopically by Fekel et al. (2000) who suggested 
the period 1259$\pm$16 d, and by Brandi et al. (2009) who suggested the value 
1290. Our \Pb~ is of course consistent with Fekel's et al. (2000) value, and 
marginally also with that of Brandi et al (2009).

It is more difficult to trace the origin of the \Pa~ period of this system.
We have already noted in  section 1 that Mira pulsations of the giant star are 
an unlike origin of the \Pa periodicity. BX Mon occupies the domain of normal M
stars in a (J-H)-(H-K) relation (Whitelock et al. 1994, Phillips 2007) and 
does not show variability in the infrared (Viotti et. al 1986). 
The origin of the \Pa~ periodicity may become clearer when we consider it as 
a linear combination of the system binary period \Pb, and the rotation period 
of the giant \Pr : 
\Pa = (1/\Pr-1/\Pb)$^{-1}$ (see Table 1).

The period \Pa~ is the length of one "day" of an
observer on the surface of the giant component of the system, whose sun is
the hot component. The giant is rotating with the
sidereal period \Pr=656 d, in the same sense as the binary revolution of the
period \Pb=1256 d. The observer will see the sun circling his or her sky in 
the retrograde direction with a period \Pa=1373 d.  If \Pr~ is the spin period 
of the giant, as is most likely the case, this is true regardless of any other 
physical assumption. 

\subsubsection{A proposed physical process}
We now suggest, in mostly qualitative terms, a physical process by which the 
ticks of this clock translate to outbursts in the LC of the system.  

If the radius of the giant star is not very small compared with the average
radius of its theoretical Roche lobe, the outer surface of it is distorted by
the tides induced by the hot component. The tidal bulge created in the giant atmosphere
rotates on its surface with the \Pa~ periodicity. This effect is well described and 
formulated in quantitative terms by Lecar, Wheeler and McKee (1976). If there is a 
region fixed on the surface of the giant, for example around a pole of a stellar 
oblique dipole magnetic field, where matter is less bound gravitationally, an 
intense mass accretion from the giant may be triggered at every passage of 
the bulge through that region. Each episode in this cyclic train of events 
liberates intense gravitational energy. It is likely to be a trigger of an 
instability outburst in the accretion disk around the WD of the system, 
liberating even larger amount of gravitational energy. Finally, the process 
may end up by large amount of hydrogen rich matter falling on the surface 
of the WD causing or enhancing already existing process of nuclear burning 
there, liberating yet another large amount of energy.

A quantitative analysis of thermodynamical and energy transfer processes that 
would show how the cyclically liberated energy is eventually appearing as 
cyclical outbursts of the optical radiation of the system is beyond 
the scope of our paper, which is concerned with the temporal behavior of the 
optical continuum emission of the system. 

However, we refer the reader to the work of Sokoloski et al. (2006) who made 
a detailed analysis of multiwavelength observations in one of the outbursts of 
the symbiotic star Z And. The optical LC of Z And shows fluctuations that are 
rather similar to those of BX Mon, in their amplitude as well as in their  
time scale.
Sokoloski et al. (2006) suggest a "combination nova" model as an interpretation 
of the spectroscopic and photometric observations, from the X-ray to the radio  
spectral regions, performed over one oscillation episode of Z And. A similar 
analysis of one light fluctuation of BX Mon require similar amount of 
simultaneous or nearly simultaneous multi-wavelength observations, which 
are not yet in existence. The present work is concerned with understanding  
the clocks that regulate this behavior.
The Z And case serves as a demonstration that qualitatively and energy-wise, 
our suggested mechanism how the diurnal cycle on the surface of the giant is 
translated to outbursts of the system is a plausible one. Naturally, much 
more work, especially in observations, is required in order to put this 
interpretation on firmer grounds.

The characteristics of the fluctuations in section B of the LC are 
similar to those in section A. They too have  the nature of outbursts 
rather than of smooth continued variability. As discussed in Section 4 
(see Fig. 3 a) and b)), even if the hot component is eclipsed by the 
giant at conjunction, the effect of it in the optical broad band 
photometry of the system is negligible. Therefore the origin of the 
intense periodic fluctuations with the binary periodicity in section B 
can hardly be attributed to the changing geometrical aspects of the 
system associated with the binary revolution.
In particular it is very unlikely that they are marks of eclipses. 
This becomes quite clear by comparing  Fig. 3 b) or Fig. 4 b) with 
Figure 1 in Skopal et al. (2000), depicting the LC of the symbiotic 
star AR Pav  in which most of the variability is indeed due to eclipses.  

The radial velocity curve of BX Mon presented by Fekel at al. (2000)
indicates that the binary orbit is quite eccentric. These authors suggest
an eccentricity of 0.45-0.55. It is therefore consistent with our
qualitative model to suggest that the fluctuations with the \Pb  
periodicity in section B of the LC have also their origin in intense mass 
accretion events in the system.  Since after the gap in the observations 
in the middle of the 20th century, these outbursts take place whenever 
the system is near its periastron passage, hence the binary periodicity.
This periodicity is hardly seen in section A  since the events of accretion 
from the bulge of the giant deplete the mass reservoir in the atmosphere 
of the giant to the extent that not much mass is left to be accreted at 
periastron passages. Once the accretion from the bulge is inhibited, 
accretion at periastron takes over.

Some support for this suggestion comes from the phase that the 
periastron passages of the system occupy in the photometric cycle. 
The vertical dashed lines in Fig. 3 b) mark the times of periastron 
passage. These are based on the date JD 2449680 of periastron passage, 
as given by Fekel et al. (2000), with all other vertical lines being 
drawn on the time axis at equal intervals of 1256 d. One can see that 
periastron passages lie on the ascending branch of the cyclic outbursts, 
as expected in our suggested scenario.  Lajoie and Sills (2011) have 
recently shown that the peak rates of mass transfer in eccentric binaries 
occur after periastron, at an orbital phase of ~0.56, independently of the 
eccentricity and mass of the stars.  Sokoloski at al. (2006) show that 
in the Z And symbiotic system, there is a delay of a few weeks between the 
time of maximum mass loss rate from the donor star and the time of the 
instability in the accretion disk triggered by it, with the associated 
liberation of gravitational energy in that process. The igniting of, 
or the extra contribution to the nuclear burning on the surface of the 
WD take place even later in time. This is why  maximum light of the 
outbursts in section B in BX Mon occurs some 350 days after periastron 
passages.     

According to this interpretation, the period change that took place between 
the years 1940 and 1972 may be a result of an inhibition that was put on mass 
accretion from the bulge of the giant. 

A process that may affect the accretion rate from the giant, inhibiting 
accretion from the bulge zone, is a  change in the intensity or the 
structure of the magnetic field of the giant. After the gap in the observations, 
the configuration of the giant global magnetic field is such that the rotating 
bulge in the atmosphere of the star does not pass anymore through an area of 
small gravitational g value. Such change in the magnetic field of the giant may 
be due to a certain magnetic dynamo cycle that operates in the outer layers of the star.

We note in this connection that in 3 other symbiotic systems, BF Cyg, YY Her
and Z And, we have also found in their historical LCs indications for the
operation of a magnetic dynamo cycle in the outer layers of the giant stars
(Leibowitz \& Formiggini 2006, Formiggini \& Leibowitz 2006, 
Leibowitz \& Formiggini 2008).
As for the idea that periodic events of intense mass transfer in  
binary systems are the cause of, or a trigger for periodic outbursts  
in the EM radiation of the system, it is of course quite an old one.  
It was discussed extensively, in particular in the context of X-ray  
binaries. See for example Murdin et al. (1980), or Stevens (1988). 

\subsection{The 7370 Periodicity}
The period  2*7370=2*\PL=14740=(1/1256-1/1373)$^{-1}$ days   
is the interval between noon time of an observer on the surface of the 
rotating giant when his sun, the hot component, is seen at a certain 
position in his sky, relative to distance stars, and the next noon 
time when he sees his sun in the same direction in his sky.
After half of this time, the sun will be seen at the observer's mid 
day at exactly the opposite direction. Thus twice every 14740/1373=10.736 
giant days, i.e. with the periodicity \PL, the maximum alignment between 
any fixed diameter line in the giant star and the radius vector from the 
center of the giant to the instantaneous L1 point of the binary system 
is reached along the direction in space of the major axis of the giant 
elliptical orbit. This can be expressed mathematically by the relation
1/\PL = .5/(1/\Pb-1/\Pa).

We note again, that if \Pr~ is the rotation period of the giant, this is 
a simple geometrical truism, regardless of any other physical consideration.
 
We can now suggest again how this clock is modulating the brightness of 
the system, as manifested by the period \PL=14740/2=7370 Earth days, 
identified in section B of the data (see Table 1). This period is the 
time interval between two successive events that we have just described, 
with the axis of a magnetic dipole field of the giant playing the role 
of the diametric axis mentioned above. In this scenario the North and 
the South poles of this field are alternating successively in getting 
closest to the L1 point during periastron passage. So while in section B, 
the timing of the intense mass transfer is controlled by the binary 
revolution through periastron passages, the magnetic field is still 
modulating to some extent the intensity of the mass transfer episodes. 
The most intense transfer events occur when the giant makes a periastron 
passage while one of its magnetic poles is closest to the L1 point of 
the system at that time. We do not pretend to know how a magnetic field 
around the L1 point affects the rate of mass transfer through this point. 
We note, however, that  the same qualitative scenario would explain the 
observed 7370 d modulation also if a magnetic field around the L1 point 
inhibits rather than intensifies mass transfer through this point.  
    
\subsection{The three remaining periods}
 
The frequency of the period \P3=431 d that appears in section B satisfies 
the equality  1/\P3=1/\Pr+1/\Pb (see Table 1). Such a beat of the \Pr~ and 
the \Pb~ periodicities will result if the intensity of a light source in the 
system that is modulated by the giant rotation period \Pr~ is modulated also 
by a cyclically varying agent with the binary period \Pb .
This would be the case if the origin of the \Pr periodicity is a stable 
or quasi stable non uniform distribution of brightness on the surface of 
the giant star, such as  the presence of large star spots. The rotation 
of the giant gives rise to the \Pr~ variability, as measure by an observer
 on Earth. This light variation is however further modulated due the system 
binary revolution, for example by obscuration of the giant by some material 
component at fixed coordinates in the binary revolving frame (e.g. gaseous 
disk around the WD). Such modulation can also be the works of the well known 
reflection effect in close binaries. The giant hemisphere facing the hot 
component is brighter than the giant other hemisphere. As the system revolves 
an observer on Earth see the brighter hemisphere periodically at the binary 
periodicity. Thus, the light from BX Mon contains a component the brightness 
of which depends on time as  I(t)=C*cos[(2*pi)/\Pr)t]*cos[(2*pi/\Pb)t],  
disregarding possible phase terms. Elementary equalities of trigonometry imply 
that the Fourier decomposition of the time series of the LC of the star should 
include a component with the frequency of the above expression. This effect was 
first discussed in details by Warner (1986). 
      
As mentioned in Section 3.1, the other two additional periods that modulate 
the optical LC of the star in section B, P1=1067 and P2=1687, can also be 
understood as beats of the two fundamental periods of the system \Pb~ and \Pr, 
or as simple linear combinations of the pair \Pb~ and \Pa~ (see Table 1). We do 
not present in this paper specific geometrical realizations of these 2 linear 
combinations. We do note, however, that various beats of the spin frequency of 
a star in an interacting binary system with the orbital frequency of the 
system is not a rare phenomenon. Appearances of such beat frequencies, or 
sidebands, are particularly prevalent in magnetic cataclysmic variables, 
as discussed by Warner (1986). Recent examples and discussions of beats 
and sidebands in LCs of interacting binaries are the works of 
Woudt et al. (2009), or of  Bloemen et al. (2010). The stellar system BX Mon 
is of course vastly different from the stars discussed in the cited examples, 
all of which belong to the family of cataclysmic (CV) stars. However the 
interplay between gravitational and magnetic effects and the varying 
aspects of the revolving binary system, which is the major cause for 
the appearance of spin-orbits beats in the LC, is common to both the 
symbiotic and the CV classes of stars. We point out, however, that 
even if no specific model can be suggested for explaining the P2 and P3 
periodicities as beats of the two fundamental frequencies in the BX Mon, 
this would not affect the analysis of the other 4 BX Mon periodicities 
that we are presenting here.  
    
\section{Summary}
We have established the existence of 2 distinct periods in the historical
LC of BX Mon. Between 1889 and 1940 i.e. section A, the brightness of 
the star fluctuated with a peak to peak amplitude of 2-3 magnitude, 
with a period of  \Pa=1373 $\pm$4 d.
This period appears in the LC with its 2nd and 4th harmonics. 
After the gap in the observations between 1940 and 1972, the LC of the star in
section B,
is again fluctuating with a peak to peak amplitude of 1-2 magnitude. This time
however, the major period is the binary period of the system \Pb=1256$\pm$16 d.

To the best of our knowledge, such an abrupt  switch between a photometric 
period of a binary system which is 10\% longer than the orbital period of the 
system, to its binary periodicity 
has no parallel in the observational study of any symbiotic star. 
It has probably neither been observed in any other binary star. 
 
In both sections of the data set we
identify periods that are various beats of the two major periodicities, in
particular in Section B we identify the \Pr=656 d period as the sidereal 
rotation period of the giant component of the BX Mon binary system. 
We suggest that the major fluctuations in the brightness of the star are due to
events of intense accretion from the giant star onto its hot companion.
Between 1889 and 1940 these events took place periodically with a period 
of 1373 d. After 1972, they are occurring in the system at the binary orbital 
frequency.    
\section*{Acknowledgments}
We acknowledge with thanks the variable star observations from the AAVSO
International Database contributed by observers worldwide and used in this
research. We also thank an anonymous referee for comments that enable  
considerable improvements in some parts of the paper.
This research is supported by ISF - Israel Science Foundation of the
Israeli Academy of Sciences. 

\label{lastpage}

\begin{thebibliography}{99}
\bibitem[\protect\citeauthoryear{}{}]{} Allen D.A., 1982, The Nature of Symbiotic Stars, eds. M. Friedjung, R. Viotti, Reidel, Dordrecht, p.27
\bibitem[\protect\citeauthoryear{}{}]{} Bessel M. S., 1983, PASP, 95, 480
\bibitem[\protect\citeauthoryear{}{}]{} Bloemen S., Marsh T.R., Steeghs D., Ostensen R.H., 2010, MNRAS, 407, 1903
\bibitem[\protect\citeauthoryear{}{}]{} Brandi E., Garcia L.G., Quiroga C., Ferrer O.E., Marchiano P., 2009, BAAA, vol. 52, 49
\bibitem[\protect\citeauthoryear{}{}]{} Dumm T., Murset U., Nussbaumer H., Schild H., Schmid H.M., Schmutz W., Shore S.N., 1998, A\&A, 336,637
\bibitem[\protect\citeauthoryear{}{}]{} Efron B., Tibshirani R.J., 1993, An Introduction to the Bootstrap, Chapman \& Hall, New York, London
\bibitem[\protect\citeauthoryear{}{}]{} Fekel F.C., Joyce R.R., Hinkle K.H., Skrutskie M.F., 2000, ApJ, 119, 1375
\bibitem[\protect\citeauthoryear{}{}]{} Formiggini L., Leibowitz E.M., 2006, MNRAS, 372, 1325 
\bibitem[\protect\citeauthoryear{}{}]{} Henden A., Munari U., 2008, Baltic Astronomy 17, 293
\bibitem[\protect\citeauthoryear{}{}]{} Iijima T., 1985, A\&A, 153, 35
\bibitem[\protect\citeauthoryear{}{}]{} Kenyon S.J., The symbiotic stars, Cambridge Univ. Press., 1986
\bibitem[\protect\citeauthoryear{}{}]{} Kenyon S.J., Webbink, R.F., 1984, ApJ, 279, 252
\bibitem[\protect\citeauthoryear{}{}]{} Lajoie C.P., Sills A., 2011, ApJ, 726, 67
\bibitem[\protect\citeauthoryear{}{}]{} Leibowitz E.M., Formiggini L., 2006, MNRAS, 366, 675 
\bibitem[\protect\citeauthoryear{}{}]{} Leibowitz E.M., Formiggini L., 2008, MNRAS, 385, 445 
\bibitem[\protect\citeauthoryear{}{}]{} Lecar M., Wheeler J.C., McKee C.F. 1976, ApJ, 205, 556
\bibitem[\protect\citeauthoryear{}{}]{} Mayall M.W., 1940, Bull.Harvard College Obs, 913, 8
\bibitem[\protect\citeauthoryear{}{}]{} Michalitsianos A.G., Kafatos M., Feibelman W.A., Hobbs R.W., 1982, ApJ, 253, 735
\bibitem[\protect\citeauthoryear{}{}]{} Munari U., Yudin B.F., Taranova O.G., Massone G., Marang F., Roberts G., Winkler H., Whitelock P., 1992, A\&AS, 93, 383 
\bibitem[\protect\citeauthoryear{}{}]{} Murdin P., Jauncey D.L., Lerche I., Nicolsonn G.D., Kaluzienski L.J., Holt S.S., Haynes, R.F., 1980, A\&A, 87,292
\bibitem[\protect\citeauthoryear{}{}]{} Phillips J.P., 2007, MNRAS, 376, 1120
\bibitem[\protect\citeauthoryear{}{}]{} Scargle J.D., 1982, ApJ , 263, 835
\bibitem[\protect\citeauthoryear{}{}]{} Skopal A., 2005, A\&A 440, 995
\bibitem[\protect\citeauthoryear{}{}]{} Skopal A., Djurasevic G., Jones A., Dreshel H., Rovithis-Livaniou H.,  Rovithis P., 2000, MNRAS, 311, 225 
\bibitem[\protect\citeauthoryear{}{}]{} Skopal A.,Vanko M., Pribulla T., Chochol D., Semkov E., Wolf M., Jones A., 2007, AN, 328, 909
\bibitem[\protect\citeauthoryear{}{}]{} Sokoloski J.L. et al., 2006, ApJ , 636, 1002 
\bibitem[\protect\citeauthoryear{}{}]{} Stevens I.R., 1988, MNRAS, 232, 199
\bibitem[\protect\citeauthoryear{}{}]{} van Belle G.T. et al., 1999, AJ, 117, 521
\bibitem[\protect\citeauthoryear{}{}]{} Viotti R., Altamore A., Ferrari-Toniolo M., Persi P., Rossi C., Rossi L., 1986, A\&A, 159, 16
\bibitem[\protect\citeauthoryear{}{}]{} Whitelock P.A., Cathpole R.M., 1983, Inf. Bull. Var. Stars No. 2296
\bibitem[\protect\citeauthoryear{}{}]{} Whitelock P.A., Menzies J., Feast M., Marang F., Carter B., Roberts G., Cathpole R.M., Chapman J., 1994, MNRAS, 267, 711 
\bibitem[\protect\citeauthoryear{}{}]{} Warner B. 1986, MNRAS, 219, 347
\bibitem[\protect\citeauthoryear{}{}]{} Woudt P.A., Warner B., Osborne J., Page K., 2009, MNRAS, 395, 2177
\bibitem[\protect\citeauthoryear{}{}]{} Zamanov R.K., Bode M.F., Melo C.H.F., Stateva I.K., Bachev R., Gomboc A., Kostantinova-Antonova R., Stoyanov K.A., 2008, MNRAS, 390, 377
\bibitem[\protect\citeauthoryear{}{}]{} Zombeck M. V., 2000 Handbook of Space Astronomy and Astrophysics, Cambridge, UK:  Cambridge Univ. Press,(2nd edition) 
\end{thebibliography}
\end{document}